\documentclass[11pt,aps,prd,nofootinbib,groupedaddress,preprintnumbers]{revtex4}
\usepackage{graphicx,epsfig}
\newcommand{\bea}{\begin{eqnarray}}
\newcommand{\beq}{\begin{equation}}
\newcommand{\eea}{\end{eqnarray}}
\newcommand{\eeq}{\end{equation}}

\usepackage{amsmath}

\DeclareMathOperator{\csch}{csch}

\DeclareMathOperator{\arctanh}{arctanh}

\newcommand{\lsim}{\raise0.3ex\hbox{$\;<$\kern-0.75em\raise-1.1ex\hbox{$\sim\;$}}}
\newcommand{\gsim}{\raise0.3ex\hbox{$\;>$\kern-0.75em\raise-1.1ex\hbox{$\sim\;$}}}

\newcommand{\unity}{{\hbox{1\kern-.8mm l}}}

\newcommand{ \overdot}{{\raise .2 ex \hbox to 0pt {\hss\bf\smash{.}\hss}}}

\begin{document}

\preprint{FTUV-14-05-37}
\preprint{IFIC-14-34}

\title{Transplanckian inflation as gravity echoes }
\author{G.~Barenboim}
\affiliation{Departament de F\'{\i}sica Te\`orica and IFIC, Universitat de 
Val\`encia-CSIC, E-46100, Burjassot, Spain.}
\author{O. Vives}
\affiliation{Departament de F\'{\i}sica Te\`orica and IFIC, Universitat de 
Val\`encia-CSIC, E-46100, Burjassot, Spain.}

\begin{abstract}
In this work, we show that, in the presence of non-minimal coupling to
gravity, it is possible to generate sizeable tensor modes in
single-field models without transplanckian field values. These
transplanckian field values apparently needed in Einstein gravity to
accommodate the experimental results may only be due to our insistence 
of imposing a minimal coupling of the inflaton field to gravity in a model with
non-minimal couplings. We present three simple
single-field models that prove that it is possible accommodate a large
tensor-to-scalar ratio without requiring transplanckian field values within the slow-roll regime.
\end{abstract}

\maketitle 

\section{Introduction}
After the recent discovery of tensor modes at BICEP2 experiment \cite{Ade:2014xna}\footnote{Throughout this work, we will assume that although the exact numbers of BICEP2  may change,
sizeable tensor modes, {\it i.e.} $ r \gtrsim .1 $ are an actual feature that will stay}, the theory of cosmological inflation \cite{inflation} can claim to be  the current (undisputed)  paradigm of early universe cosmology. Inflation cannot only solve most of the problems of the Standard Big Bang Model, but it offers the only available  explanation for the origin of the large-scale structure of the universe based on causal physics.  Even more, cosmological inflation is a predictive theory.  It calls for an almost scale invariant spectrum of curvature perturbations which anticipates  the characteristic oscillations in the angular power spectrum of cosmic microwave anisotropy maps,  observed with high accuracy by WMAP \cite{Bennett:2012zja} and Plank \cite{Ade:2013uln}.

Inflation is simply  the assumption that there was a short epoch in the very early universe where  the scale factor (space)  grew at an accelerated pace, typically in an exponential way. Such an accelerated expansion flattens out and widens up a microscopic size of space,  solving the longstanding ``size problem'' of standard cosmology. Not only that, the accelerated expansion decreases the contribution of any pre-existing curvature to the total energy budget of the universe and therefore turns the spatially flat universe into a local attractor in initial-condition space, solving this way yet another cosmological puzzle, the ``flatness problem''.
Unfortunately, inflation comes at a cost. Successful models of inflation,  {\it i.e.} successful inflationary potentials require unusual features:  the potentials have to be extremely flat so that enough inflation is produced to actually solve the above-mentioned issues, and observations seem to require the inflaton field to travel over transplanckian distances in field space.  In fact, following an argument due to Lyth \cite{Lyth:1996im} \footnote{This bound has been generalized in the context of effective field theories of inflation in Ref.~\cite{Baumann:2011ws}. However, for the sake of this work the original bound still holds.} we have,
\beq
\frac{\Delta \theta}{M_{\rm Pl}} \gtrsim 5.8  \left(\frac{N_e}{50}\right) \left(\frac{r}{0.2}\right)^{1/2}
\eeq
with $\Delta \theta$ the variation of the field during inflation, $r \simeq 13.8 ~\epsilon$ the tensor-to-scalar ratio with $\epsilon$ the usual slow-roll parameter, $N_e$ the number of e-folds of inflation since the relevant scales left the horizon till the end of inflation and $M_{\rm Pl} = (16 \pi G)^{-1/2}$. Therefore, the value of $r = 0.2^{+0.07}_{-0.05}$ measured at BICEP2 implies transplanckian values for the inflaton field, 
$\Delta \theta/M_{\rm Pl} \gtrsim 5.8$. Fortunately, large  (transplanckian) field values  do  not necessarily involve large (transplanckian) energies, which is the reason why  transplanckian field values are not total anathema. In fact, transplanckian field values have been the norm rather than the exception in the inflationary game \cite{Chialva:2014rla,Antusch:2014cpa,Kehagias:2014wza,Mazumdar:2010sa}. There is (almost) no single-field inflationary model which can be kept below Planck scale all the way.
Yet another problem which has not been devoted enough attention to is the fact that the energy scale of inflation and the Planck scale are not that far from each other and therefore it is easy to imagine that corrections to Planck scale physics are bounded to play a role.
Whether this role is significant or not is clearly a debatable issue. 
Going back to the transplanckian field values, one of the reasons why it is safe to entertain transplanckian field values (once checked that the observables are well behaved) is that a field  is, after all, a ``dummy'' variable, {\it i.e.} it is ``per-se'' meaningless. Just a field redefinition will turn its value into the desired domain at no expense, all the observables will remain invariant.  Nevertheless, field redefinitions may be gratis observable-wise, but they are not innocent. They will surface somewhere else: in a change of the kinetic terms, the couplings in the potential, etc. In the same way that the mass matrix in the quark sector can be made real, but then the removed (physical) phase will show up in the charged and neutral current interactions, a field redefinition to turn the inflationary field subplanckian may end up shedding light on the shape of gravity close to the Planck scale.

In this work, we conjecture about the possibility that the trasplanckian field values arising in single-field inflationary models may be due to the fact that we are ``forcing'' our model to have Einstein gravity.  We will show that well-behaved and subplanckian modified gravity, as non-minimally coupled scalar fields and/or scalar tensor theories, can become transplanckian once forced to behave as minimally coupled scalar field theories in Einstein gravity. Therefore, the observed tensor-to-scalar ratio can be obtained in
single-field inflation models, in the presence of non-minimal couplings
to gravity, working always in the subplanckian regime and in the slow-roll approximation.

This work is organised as follows. We begin with a basic review of
models with non-minimal coupling to gravity and recall the use of
conformal transformations to go from the Jordan frame (with
non-minimal coupling to gravity) to the Einstein frame in section
\ref{sec:jordan}. In section \ref{sec:models}, we present several
realistic examples showing the effect of conformal transformations in
the field values, making subplanckian field values in the Jordan
frame transplanckian in the Einstein frame. Finally, results and
conclusions are summarised in section \ref{sec:conclusions}.

\section{Inflation in theories with non-minimal couplings to gravity.}
\label{sec:jordan}
We start from a general theory with gravity coupled to a single scalar field that will play the role of the inflaton. The action in the Jordan frame, with non-minimal coupling to gravity and assuming canonical kinetic terms\footnote{The introduction of a non-canonical kinetic term will complicate 
unnecessarily the theory and is not needed to make our point clear}, would be,
\begin{eqnarray}
S = - \int d^ 4 x ~\sqrt{-g} \left[\frac{k^ 2}{4} D(\theta)~ R - \frac{1}{2} ~g^ {\mu \nu}~ \partial_\mu \theta \partial_\nu \theta + V(\theta) \right]
\end{eqnarray} 
where $R$ is the scalar curvature and $\theta $ our scalar field\footnote{This action becomes non-renormalizable, once the field is above the cutt-off scale in the Einstein frame, a fact that may {\it per-se} be signaling the need to introduce a non-minimal coupling to gravity as the true driver of inflation as higher correction are always kept under control in this frame} .
In the absence of any other sources of matter, and specialising for the case of a Friedmann-Robertson-Walker metric, $g_{\mu\nu} = \mbox{diag} \{ 1, -a(t)^2, -a(t)^2,-a(t)^2 \}$ 
the equations for the Hubble rate and the $\theta$ field become,
\begin{eqnarray}
\label{einstein}
&D(\theta)~ H^2 = \frac{\dot{\theta}^2}{3 k^2 } + \frac{2 V (\theta)}{ 3 k^2} - \dot{D}(\theta)~ H& \\
&\ddot{\theta} + 3 H \dot{\theta} + \frac{k^2}{4}~D'(\theta)~R +V'({\theta})=0 &\, ,\nonumber
\end{eqnarray} 
with $k^2=M_{\rm Pl}^2/(4 \pi)$, from where it is straightforward to obtain,
\begin{eqnarray}
\label{dotH}
\dot{H} = -\frac{\dot{\theta}^2}{k^2 D} + \frac{\dot{D}(\theta)}{D(\theta)}~ \frac{H}{2} - \frac{\ddot{D}(\theta)}{2 D(\theta)}\, . 
\end{eqnarray} 

Due to the addition of the extra source for perturbations we have introduced, $D(\theta)$, we need to include two more slow-roll parameters as compared to the standard case \footnote{Throughout this work we assume the slow-roll regime. In principle the field could fast roll and in this case, quantum corrections can become sizeable and the slow-roll solution might stop being an attractor}. The scalar-type perturbations will be affected by both of them, although only one ($ \epsilon_3 $) will be relevant for the tensor perturbations \cite{Hwang:1996xh,Noh:2001ia},
\begin{eqnarray}
\label{slow-roll}
\epsilon_1 &=& \frac{\dot{H}}{H^2} =\frac{H' \dot{\theta}}{H^2} \\  
\epsilon_2 &=& \frac{\ddot{\theta}}{H \dot{\theta}} \\ 
\epsilon_3 &=& \frac{1}{2} \frac{\dot{D}}{H D} =  \frac{1}{2} \frac{D' \dot{\theta}}{H D} \\
\epsilon_4 &=& \frac{1}{2} \frac{\dot{E}}{H E} =  \frac{1}{2} \frac{E' \dot{\theta}}{H E} \, ,
\end{eqnarray}
with $E=  3k^2 (D')^2/2 + D$. Assuming $\dot{\epsilon}_i =0$ 
and to linear order in the slow-roll parameters
\begin{eqnarray}
n_s &=& 1+ 2\left( 2\epsilon_1 - \epsilon_2 + \epsilon_3 - \epsilon_4 \right) \\
n_T &=& 2\left( \epsilon_1 -  \epsilon_3  \right) \\
r &=& 13.8 \mid \epsilon_1 - \epsilon_3 \mid \label{rjordan}
\end{eqnarray}

As it is well-known, this model, as any non-standard theory of gravity, can be mapped into a standard theory of gravity at the expense of having a more complicated matter sector by a conformal transformation.  Such a transformation is not just a coordinate redefinition (being general relativity a covariant theory, a coordinate redefinition would become trivial) rather, it is a transformation that mixes up the matter and gravitational degrees of freedom\footnote{Every single choice of field (and metric) definition among the family of transformations goes under the name of {\bf frame} and obviously the frame where gravity takes the form of Einstein's theory is called Einstein's frame.}.

The mapping we are alluding to, takes the original metric $g_{\mu \nu}$ into a new metric $ \tilde{g}_{\mu \nu}$ according to, 
\begin{equation}
\tilde{g}_{\mu \nu} = e^{2 \omega} g_{\mu \nu}\, ,
\end{equation}  
with $ e^{2 \omega} = D(\theta)$. 

The Hubble rate transforms as,
\beq
\label{Htrans}
\tilde H = \frac{ H + \dot D(\theta)/ (2 D(\theta))}{\sqrt{D(\theta)}}\,,
\eeq
with $\dot D = \partial D /d t$, and the canonically normalised field replacing $\theta$ in the Einstein frame is, 
\begin{equation}
\label{fieldtrans}
\phi(\theta) = \pm \int \sqrt{\frac{3}{2} \left(\frac{D'(\theta)}{D(\theta)}\right)^ 2 + \frac{2}{k^{2} D(\theta)}}~d \theta\,.
\end{equation} 
In terms of this rescaled field the action takes the form,
\begin{eqnarray}
S = - \int d^ 4 x ~\sqrt{-\tilde g} \left[\frac{k^ 2}{4} \tilde R - \frac{1}{2} ~\tilde g^ {\mu \nu}~ \partial_\mu \phi \partial_\nu \phi + \tilde V(\phi) \right] \, ,
\end{eqnarray}
with $\tilde V(\phi) = V(\phi(\theta))/D(\theta)^2$. 

It is trivial to show that the slow-roll parameters in both frames are related as,
\begin{eqnarray}
\label{Einst-roll}
\tilde \epsilon &=& \frac{k^2}{4}\left(\frac{\tilde H'}{\tilde H}\right)^2 = \epsilon_1 -\epsilon_3 \\  
\tilde \eta &=& \frac{k^2}{4}\frac{\tilde H''}{\tilde H}= \epsilon_2 - 3 \epsilon_3 + \epsilon_4 \\
\tilde{n}_s &=&1+ 2\left( 2\tilde{\epsilon} -   \tilde{\eta} \right) = 1+ 2\left( 2\epsilon_1 - \epsilon_2 + \epsilon_3 - \epsilon_4 \right) = n_s \\
\tilde{n}_T &=& 2\tilde{\epsilon} =  2\left( \epsilon_1 -  \epsilon_3  \right)= n_T \\
\tilde{r} &=& 13.8 \mid \tilde{\epsilon} \mid = 13.8 \mid \epsilon_1 - \epsilon_3 \mid =r \label{rEinstein}
\end{eqnarray}
Leaving all the observables invariant, as it is obvious given the fact that changing from one frame to another one does not correspond to a change in the physics. However as the conformal transformation changes the space-time curvature (and also the scalar/matter field) phenomena that appear to be due to gravity in one frame may appear to be originated in the scalar sector in another. 
Besides, it is easy to see that as a result of the fact that the inflaton field in Einstein and Jordan frames are related in a highly non-trivial way,  it can be expected that subplanckian values in a given frame, may correspond to transplanckian values in the second frame.  

The analysis in this paper is done in the framework of and effective
field theory neglecting terms suppressed by the cutoff scale, which in
the Einstein frame is $M_{Pl}$. However, as pointed out recently by
Hertzberg in Ref.~\cite{Hertzberg:2010dc}, in the presence of non-minimal couplings to
gravity the validy regime of the effective theory may change. As shown in
this work, in single field models, the relevant cutoff scale, in the
gravitational and kinetic sector, is still the Planck mass. Regarding
the potential interactions the situation is model dependent and we
will check it in a case by case basis.

\section{Single-field non-minimal models of inflation}
\label{sec:models}

As shown in the previous section, the field values in two different
frames are correlated by an non-trivial function in a rather complicated
way. Here, we will show that is possible and, in fact, quite natural and easy to
find realistic examples of theories with non-minimal coupling to
gravity, modified gravity or scalar tensor theories, that have
transplanckian field values if we insist on imposing a minimal
coupling to gravity, but are always subplanckian in their ``natural''
frame. This clearly does not imply that any conformal
transformation will turn transplanckian minimally coupled scalar
fields into the subplanckian regime once non-minimally coupled to
gravity or once allowed to live in a modified-gravity framework but, in our
scheme, observations would select a subclass of conformal transformations.

\subsection{Monomial Potentials}
\label{monomial}
As a first toy-model, we assume that the potential in the Einstein frame is exactly given by
the well-known potential $V(\phi) = \lambda \phi^4$ in the Einstein frame. 
In the slow-roll regime $\dot \phi^2 \ll V(\phi)$, using the Eqs. of motion, we have,
\begin{eqnarray}
H& =& \sqrt{\frac{2 \lambda}{3}} \frac{\phi^2}{k} \\
\dot \phi &=& - 2 \sqrt{\frac{2 \lambda k^2}{3}} \phi 
\end{eqnarray}
and therefore, we have,
\begin{eqnarray}
\epsilon &=& \frac{\dot H}{H^2} =  \frac{H' \dot \phi}{H^2} =  - \frac{4 k^2 }{\phi^2} \\
\eta &=& \frac{\ddot \phi}{H \dot \phi} = - \frac{2 k^2 }{\phi^2} \\
n_s &=& 1 + 4 \epsilon - 2 \eta = 1 - \frac{12 k^2 }{\phi^2} 
\end{eqnarray}
Now, the number of e-foldings fixes the value of the field at which the scales of interest at present left the horizon,
\begin{equation}
N = \int H dt = - \left.\frac{\phi^2}{4 k^2}\right|^{\phi_f}_{\phi_i} \simeq \frac{\phi_i^2}{4 k^2}  
\end{equation}
Using $N \simeq 62$ we need $\phi_i \simeq 11 \times k \simeq 3.1 \times M_{\rm Pl}$, and we obtain $n_s= 1 - 3 /N \simeq 0.95$ and $r \simeq 13 ~\epsilon = 13/N \simeq 0.21$. Therefore, we see that this potential would be able to reproduce approximately the observed values for the spectral index and the tensor-to-scalar ratio, but only if the field values during inflation are well above the Planck mass.

However, if our field makes excursions well-beyond the Planck scale, or gets very close to it, we can expect gravitational corrections
to come into play and to be very relevant. For example, higher-order curvature invariants could appear, and it is then 
natural to consider also non-minimal couplings of the inflaton to gravity. 

Now, let's assume that our inflaton field, $\theta$, has a
non-minimal coupling to gravity \footnote{In this case, although the
potential is exactly quartic in the Einstein frame, the effects of this 
non-minimal coupling would still appear in this frame as higher order terms in the action that we have not considered} of the form  $D(\theta)=(1 - \theta^2/( 3 k^2))$. The Einstein equations and the $\theta$ equations of motion in this frame, the Jordan frame, are,
\begin{eqnarray}
&D(\theta)~ \tilde H^2 = \frac{\dot{\theta}^2}{3 k^2} + \frac{2 \tilde V (\theta)}{ 3 k^2} - \dot{D}(\theta)~ \tilde H& \\
&\ddot{\theta} + 3 \tilde H \dot{\theta} + \frac{k^2}{4}~D'(\theta)~R + \tilde V'({\theta})=0 &\, ,\nonumber
\end{eqnarray} 
with $\tilde H$ the Hubble rate in the Jordan frame related to the Hubble rate in the Einstein frame by Eq.~(\ref{Htrans}). Then, the potential in the Jordan frame is,
\beq
\label{thetapot}
\tilde V(\theta) = \frac{V\left(\phi(\theta)\right)}{(D(\theta))^2} = \lambda \frac{ \left(\phi(\theta)\right)^4}{\left(1 -\theta^2/(3 k^2) \right)^2}
\eeq
Using this potential we can obtain, folowing \cite{Hertzberg:2010dc},
the Jordan-frame cutoff scale wich signals the validity regime of the
effective theory after taking into account the quantum corrections 
incorporated a la Coleman-Weimberg. In this case the cutoff is even larger 
than the Einstein frame one, and it is given by $\Lambda = 6 M_{Pl}$.

The fields in the Jordan and Einstein frames, are related by Eq.~(\ref{fieldtrans}), that in this case can be integrated analytically,
\begin{equation}
\label{fieldredef}
\phi(\theta) = 2 \sqrt{6 \pi}~k \arctanh\left[\frac{\theta}{\sqrt{3}~ k}\right] \,,
\end{equation}
or, 
\begin{equation}
\theta(\phi) = \sqrt{3}~k \tanh\left[\frac{\phi}{2\sqrt{6\pi}~ k}\right] \,.
\end{equation}

Therefore, we can see clearly that in the Jordan frame, the field $\theta$ is always subplanckian, $\theta \leq \sqrt{3/(4\pi)}~ M_{\rm Pl} \simeq 0.489~ M_{\rm Pl}$, and when $\phi \simeq 3.1~ M_{\rm Pl}$ the Jordan field $\theta \simeq 0.42~ M_{\rm Pl}$\footnote{Here, we consider only the leading term in the potential. It is clear that higher-order operators could play a role. However, it is always possible to deal with them through a symmetry. For example, in this case, we could impose a $Z_4$ symmetry.}. It is a trivial exercise to show that using Eqs~(\ref{slow-roll}--\ref{rjordan}) and (\ref{Einst-roll}--\ref{rEinstein}),  and despite the fact the slow-roll parameters are different in both actions, two non-vanishing in one case corresponding to the usual slow-roll parameters, and four in the Jordan frame, all the observables are identical. Moreover, even in the Jordan frame, the potential is approximately quartic in $\theta$ at low field values, $\theta/(\sqrt{3} k) \ll 1$, as can be seen from 
Eqs.~(\ref{thetapot}) and (\ref{fieldredef}). Therefore, already by the end of inflation,
 both theories are nearly indistinguishable\footnote{An interesting possibility would be to have  the same non-minimal coupling to gravity to play a role in the current accelerated expansion.}.

We can repeat the same exercise starting from a quadratic potential $V(\phi) = \mu^2 \phi^2$. In this case, we obtain $\epsilon = 1/(2 N_e)$, $n_s = 1 - 2/N_e$ and $r=6.9/N_e$, with $N_e = \phi_i^2/(2 k^2)$ the number of e-folds needed for inflation to solve the cosmological problems, which require $\phi_i\simeq 2.2 \times M_{\rm Pl}$. So, again we need transplanckian values for the field in the Einstein frame (less transplanckian than in the previous case due to the flatter potential), but clearly, assuming the same non-minimal coupling to gravity, the relation between the Jordan and Einstein-frame fields is the same as in Eq.~(\ref{fieldredef}), and therefore, as in the $V=\lambda \phi^4$ case, transplanckian field values become subplanckian, $\theta_i \simeq 0.35 \times M_{\rm Pl}$, once allowed to couple to the curvature.

\subsection{Generic scalar-tensor theories}
\label{generic}
In the previous model, we have specified the potential in the Einstein frame and the transformation to the Jordan frame, and we have seen that the transplanckian values of the field may be simply due to our attempt to write in Einstein form a theory that has a non-minimal coupling to gravity.

Here we will use a different strategy, we will start from an ansatz that guarantees inflation in the Jordan frame and obtain the potential and the non-minimal coupling to gravity from there.

We start from the requirement that space inflates exponentially with the inflaton field being 
responsible for it, $ a = \exp{\left( -\theta/b \right)}$ and therefore $H = \dot a /a = - \dot \theta/ b$ \cite{Barenboim:2007bu}.
 
This ansatz establishes also the number of e-foldings in this scenario, which is given by
\beq 
N_e = \int H dt = - \int \frac{\dot \theta}{b} dt = \frac{-1}{b} \int d\theta = \frac{1}{b} \left(\theta_i - \theta_f\right) \simeq \frac{\theta_i}{b}\,,
\eeq
where the fact that the scalar field is rolling down ($\theta$ is decreasing)
becomes transparent  and we have chosen $\theta_f =0$ at the end of inflation for simplicity. 
  
Using Eqs.~(\ref{einstein}) and (\ref{dotH}), with $H=-\dot \theta/b$ and $\dot f = f' \dot \theta$, we 
can now obtain the relation between the Hubble rate and the coupling to gravity that will sustain the
exponential period of expansion we are longing to have,
\beq
\frac{H'}{H} = \frac{ 2 b/k^2 +b D''  + D'}{2 D - b D'}
\eeq

The following step is clear, we need to choose either a coupling to gravity (as we did in the previous section) or a Hubble rate,
and then obtain the other one via this second order differential equation. 
In this section, we are going to choose the form of the Hubble rate and, from there, obtain the non-minimal coupling to gravity. For simplicity we want to obtain analytic expressions for this coupling, and then not many choices for $H'/H$ are possible. 

Unfortunately the most natural and easiest choice, 
$H'/H \simeq 0$, which gives,
\beq 
D (\theta) = -\frac{2 b \theta}{k^2} - A~ b e^{- \theta /b} + B\,,
\eeq
where $A$ and $B$ are the two integration constants, with  $B$ dimensionless and $[A] = 1/[b]= 1/[\theta]$, fails phenomenologically. It gives a red spectral index,
($n_s > 1 $) and therefore must be abandoned.
 
The next choice (in order of simplicity) would be $H'/H = -1/ M$. We can still solve exactly the equation
for the coupling to curvature although the solution is not as simple as before,
\bea 
\label{Dtheta}
D (\theta) =  -\alpha  \frac{M^2 }{k^2} +e^{\frac{\left(\alpha -1 +\sqrt{\alpha ^2-10\alpha  +1 }\right) \theta}{2 \alpha M}} (\alpha \frac{M^2 }{k^2}+ 1 -B) +e^{\frac{\left(\alpha  -1-\sqrt{\alpha ^2-10 \alpha  +1}\right) \theta}{2 \alpha M }} B\,,
\eea
with $B$ a dimensionless integration constant, which in the following we fix at $B=0$ for convenience, and $\alpha = b/M$. We have fixed the second integration constant requiring that $D(\theta_f=0)=1$, so that at the end of inflation we naturally land in an
Einstein gravity regime. Despite its rather complicated form, we will see that, for the set
of parameters needed to produce the correct inflation phenomenology, the behaviour of $D(\theta)$ is
not as sophisticated as it can appear by looking at its full expression.
\begin{figure}
\centerline{\epsfxsize 4. truein \epsfbox {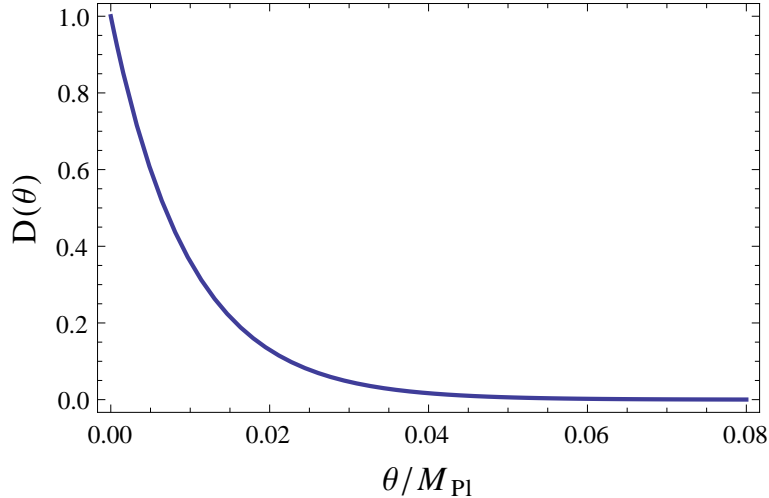} } 
\caption
{Coupling of the scalar field to the Ricci scalar curvature $R$. The figure is produced for the slow-roll regime, $\alpha = 0.007$ and taking $B=0$. The value of $M$ is irrelevant as long as $M \lesssim 0.1 M_{\rm Pl}$. The field value is given in units of $M_{\rm Pl}$
\label{dfun}}
\end{figure}
The number of e-foldings in this scenario can be written in terms of the new mass scale $M$ and
the parameter $\alpha$ as $N_e \simeq \theta_i/b = \theta_i/(\alpha M)$. 

For the slow-roll parameters defined in Eqs.~(\ref{slow-roll}), we have, 
\begin{eqnarray}
\label{sroll2}
\epsilon_1 &=& \frac{H' \dot{\theta}}{H^2} = -b  \frac{H'}{H} = -\alpha \\  
\epsilon_2 &=& \frac{\ddot{\theta}}{H \dot{\theta}} =  \frac{(\dot{\theta})'}{H } = - b  \frac{H'}{H} =  -\alpha\\ 
\epsilon_3 &=&\frac{1}{2} \frac{D' \dot{\theta}}{H D} = -\frac{b}{2} \frac{D'}{D} \simeq  - \frac{\left( 1 + \frac{M^2}{k^2} \alpha\right) \cdot \omega(\alpha)e^{-\frac{N_e}{2}\cdot \omega(\alpha)} }{4\left(\frac{M^2}{k^2} \alpha - \left( 1 + \frac{M^2}{k^2} \alpha\right) e^{-\frac{N_e}{2}\cdot\omega(\alpha)}  \right)}\\
\epsilon_4 &=& \frac{1}{2} \frac{E' \dot{\theta}}{H E} = -b  \frac{2 D' + 3 k^2 D' D''}{4D + 3 k^2 (D')^2} \simeq -\left(1+ \frac{M^2}{k^2} \alpha\right)~ e^{-\frac{N_e}{2}\cdot \omega(\alpha)} \cdot \\
&& \frac{\frac{M^2}{k^2} \alpha^2 \cdot \omega(\alpha)  + 3 \left(1 + \frac{M^2}{k^2} \alpha \right) \left((1-4\alpha +\alpha^ 2) \cdot \omega(\alpha) - 4 \alpha (1-\alpha)\right)~ e^{-\frac{N_e}{2}\cdot \omega(\alpha)}}{4 \frac{M^4}{k^4} \alpha^3 -4  \frac{M^2}{k^2} \alpha^2 \left(1 + \frac{M^2}{k^2} \alpha \right)~ e^{-\frac{N_e}{2}\cdot \omega(\alpha)}+ 3  \left(1 + \frac{M^2}{k^2} \alpha \right)^2((1-\alpha) \cdot \omega(\alpha) - 4 \alpha) ~ e^{-N_e\cdot \omega(\alpha)}} \, ,\nonumber
\end{eqnarray}
with $\omega(\alpha)=(1 - \alpha -\sqrt{1 - 10 \alpha +\alpha^2})$ and always taking $B=0$ in Eq.~(\ref{Dtheta}).

These expressions are simplified in the limit $\alpha \ll 1$, 
corresponding to the slow-roll regime,
\begin{eqnarray}
\label{sroll3}
\epsilon_3 &\simeq&  - \alpha~ e^{-2 N_e\alpha} \frac{(1 +3 \alpha) }{\left(\frac{M^2}{k^2} \alpha - e^{- 2 N_e\alpha}  \right)} \\
\epsilon_4 &\simeq& -\alpha ~ e^{-2 N_e\alpha} \frac{\frac{M^2}{k^2} (1+ 3 \alpha) + 12 ~ e^{- 2 N_e\alpha}}{\frac{M^4}{k^4} \alpha - \frac{M^2}{k^2}~ e^{- 2 N_e \alpha}- 6~ e^{- 4 N_e \alpha} } \, .
\end{eqnarray}
The usual slow-roll parameters, in the interesting region $\frac{M^2}{k^2}< e^{- 2 N_e \alpha}$ ( {\it i.e.} basically $120~ \alpha \sim O(1)$), become,
\bea
\tilde \epsilon& =& \epsilon_1 - \epsilon_3 \simeq -2  \alpha \\
\tilde \eta &=& \epsilon_2 + \epsilon_4 - 3 \epsilon_3 \simeq - 2 \alpha  \\
n_s &=& 1 + 4 \tilde \epsilon - 2 \tilde \eta \simeq 1 -4  \alpha\\
r &=& 13.8 \left| \tilde \epsilon \right| \simeq 27.6 \alpha\,.
\eea
And, for $\alpha = 0.007$, we obtain $r\simeq 0.19$ and $n_s\simeq 0.97$. As explained before, the value of $M$ is irrelevant for these observables as long as $M \lesssim 0.1 M_{\rm Pl}$. 

\begin{figure} 
\centerline{\epsfxsize 3.3 truein \epsfbox {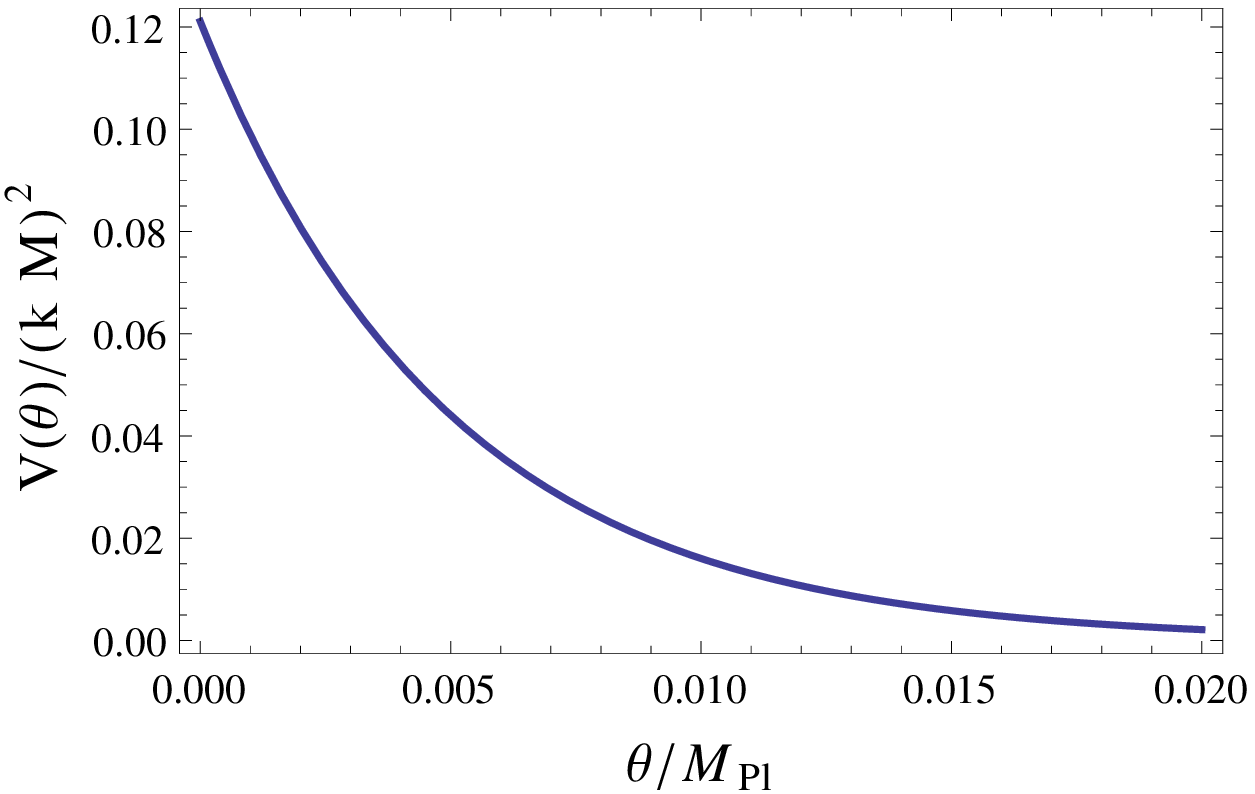} \epsfxsize 3.3 truein \epsfbox{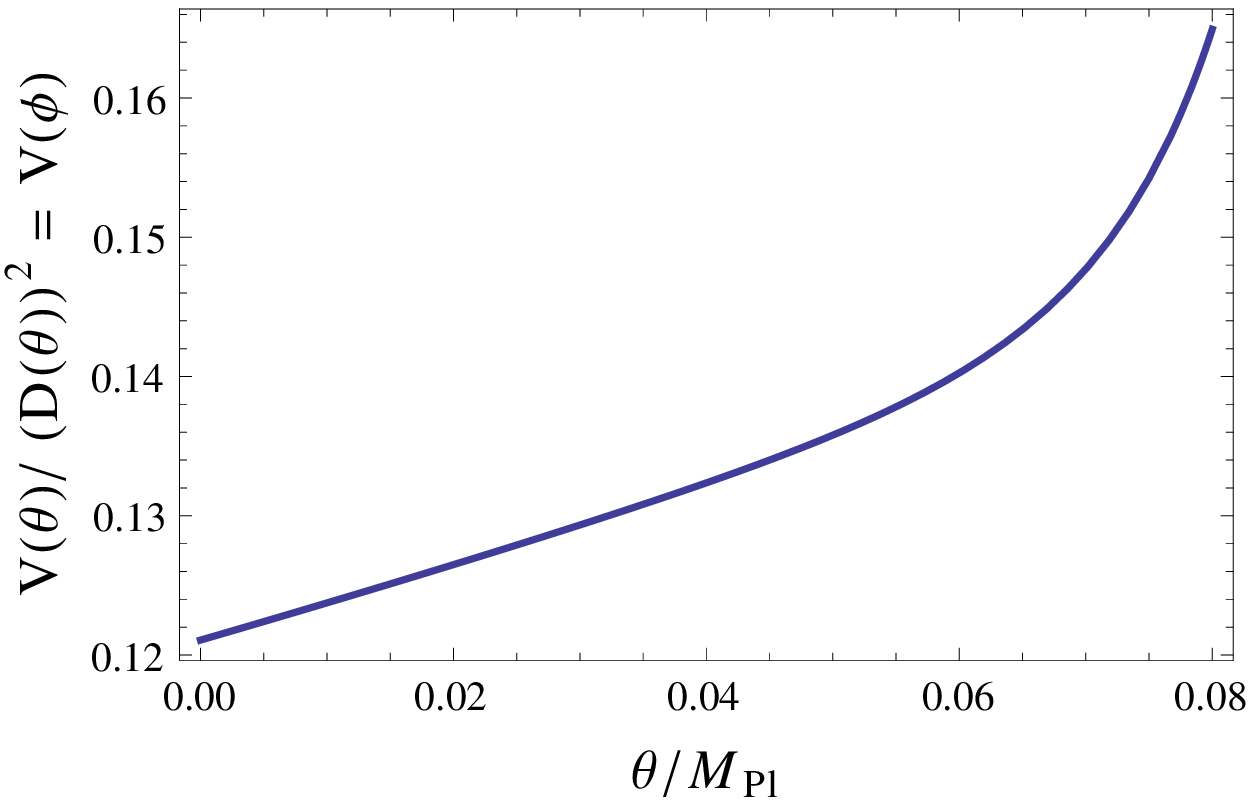} } 
\caption
{Potential of the scalar field in the Jordan frame  in terms of the non-minimally
coupled field (left). The potential in the Einstein frame in terms of the non-minimally coupled
variable is shown in the right-side panel.  The figures are produced for the slow
roll regime, $\alpha = 0.007$ and taking $B=0$. The value of $M$ is irrelevant as long as $M \lesssim 0.1 M_{\rm Pl}$. The field value is given in units of $M_{\rm Pl}$. 
\label{potential}}
\end{figure}

\begin{figure} 
\centerline{\epsfxsize 4. truein \epsfbox {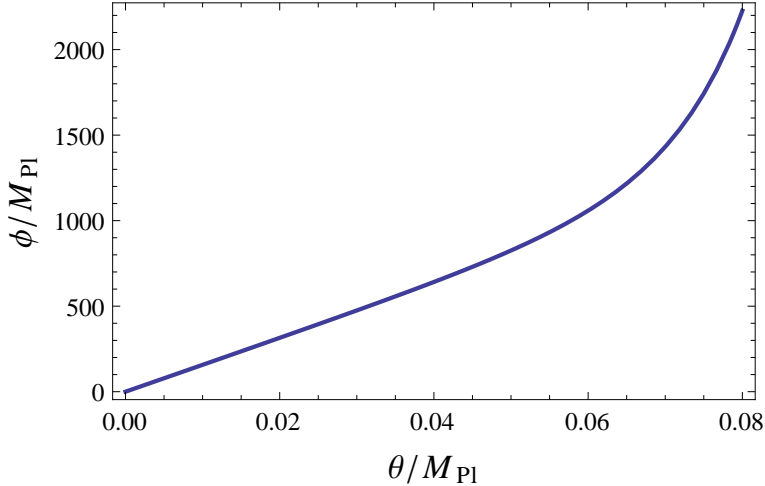} } 
\caption
{Minimally coupled scalar field in terms of the non-minimally
coupled field. The figure is produced for the slow
roll regime, $\alpha = 0.007$, $M=0.02$ and taking $B=0$. 
Both field values are given in units of $M_{\rm Pl}$. From the figure it becomes apparent
that while the field in the Einstein frame is transplanckian, the one in the Jordan frame is not.
\label{field}}
\end{figure}

As before, we can get the potential, which has a rather baroque expression in full form, although it is basically an exponential potential $ e^{-\frac{2 \theta }{M}}$,
\bea
V(\theta)&=&\frac{3 \; k^2}{4} e^{- \frac{2 \theta }{M}}\left[\left(1+\frac{M^2  }{k^2} \alpha\right)( 2 + \omega(\alpha))~ e^{-\frac{- \theta \cdot \omega (\alpha ) }{2 M \alpha }} -\frac{2}{3}  \frac{M^2}{k^2}\alpha (3 + \alpha) \right] \,,
\eea 
as can be seen in the left-side plot in Fig.~\ref{potential}. Here, we see that the potential in the $\theta$ field is decreasing and seems not able to produce inflation. However, in the Jordan frame, we must take also into account the effects of the non-minimal coupling to gravity, and then the corresponding potential in the Einstein frame becomes much more attractive. This is shown in the right-side plot in Fig.~\ref{potential}, where we plot the potential in the Einstein frame as a function of the $\theta$ field (the Einstein potential in terms of the $\phi$ field is obtained changing variables using Fig.~\ref{field}).    
In this case, the exponential potential is well-behaved and the
quantum corrections at one-loop are again small in the region of
interest for inflation. The inflaton mass is always smaller than
$M_{Plank}$. Thus as shown in Ref.~\cite{Hertzberg:2010dc} the cutoff
is still $M_{Plank}$ and there are no UV issues below that scale.

Courtesy of the sophisticated form of $D(\theta )$, it is clear that the conformal
transformation, which will turn the $\theta $ field into a more familiar minimally 
coupled scalar field enjoying Einstein gravity, cannot be carried out analytically.
In Fig.~\ref{field},  we show the relation between both fields, where, as before, our point becomes
transparent: the non-minimally coupled field is subplanckian at $\theta_i= \alpha M N_e \simeq 0.01~ M_{\rm Pl}$, while the minimally coupled one is not, $\phi_i \simeq 140~ M_{\rm Pl}$.
At this point, it is important to stress that, in this case, we have not engineered the
coupling to gravity in order to support our point. Instead, we have only asked our scale factor
to sustain an inflationary period and looked for the simplest possible choice allowing us to 
solve analytically the second order differential equation which relates the Hubble rate to 
the curvature coupling. In this context, the emergence of a subplanckian field value in the
Jordan frame cannot be considered the result of a fine-tuning.

\subsection{f(R) gravity models}
\label{fR}
\begin{figure}
\centerline{\epsfxsize 4. truein \epsfbox {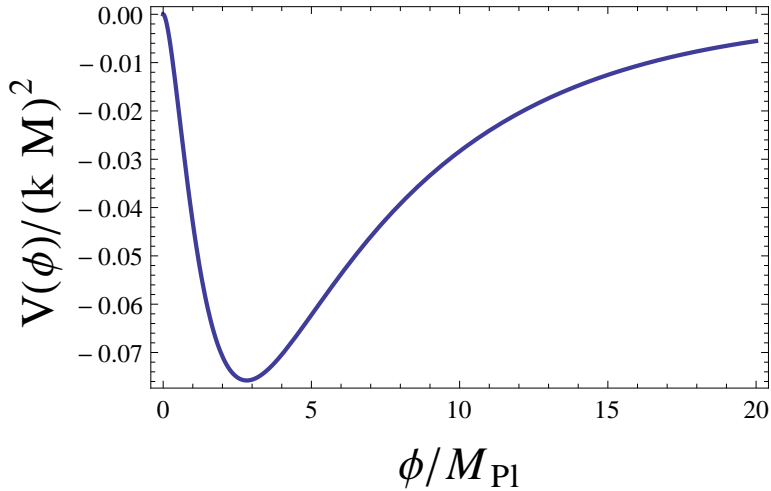} } 
\caption
{Potential for the scalar field introduced in the conformal transformation. From the figure it is obvious that the potential does have the right shape to inflate. However transplanckian masses are involved.
\label{fr}}
\end{figure}

Unlike the previous cases, where inflation was given by the interaction between the matter
sector and the modified gravity sector, 
to finish  we will consider the case where inflation is entirely nourished by gravity \cite{Sotiriou:2008rp},
\beq 
L = \frac{k^2}{4} f(R) \,.
\eeq
In this case, equations for the background become,
\bea
H^2 &=& \frac{1}{3 F(R)} \left( \frac{R F(R) -f(R)}{2} - 3 H \dot{F}(R) \right)\,,
\nonumber \\
\dot{H}& =& - \frac{1}{2 F(R)} \left( \ddot{F}(R) - H\dot{F}(R) \right)\,,
\eea
where $F(R) = \partial f(R)/\partial R$ and $R = - 6 (2 H^2 + \dot{H}) $.
Face value, this case is quite distant from the previous ones, as now there is no conformal 
transformation capable of driving us to the Einstein frame. However, once we depart
in a non-trivial way from the standard gravity, the field equations for $R$ become
higher-order effectively, signalling the  presence of additional degrees of freedom.
This feature can be taken care by the introduction of an auxiliary scalar field and going, as an intermediate step, through a Brans-Dicke form of our model \cite{Sotiriou:2008rp}.
Then, in a similar way as in the previous cases, a conformal transformation will take us 
away from our $f(R)$ gravity to the kingdom of Einstein gravity plus a minimally coupled scalar field with a specific potential. In the case we are studying,  
under a conformal transformation, the metric is redefined as,
\beq
\hat{g}_{ab} = \Omega^2 g_{ab}\,,
\eeq
where $\Omega $ is a spacetime position-dependent factor and is defined to be, 
\beq
\Omega^2 = F(R) = \exp{\left(\sqrt{\frac{2}{3 k^2}} \phi\right)}\,,
\eeq
and $\phi$ is the new dynamical variable we have obtained after conformal transformation of the Brans-Dicke auxiliary field,
\beq
\phi = \sqrt{\frac{3 k^2}{2}} \ln F(R)\,.
\eeq
The Lagrangian in terms of the $\phi$ field is given by, 
\beq
L = - \left( \frac{k^2}{4}\hat{R}  - \frac{1}{2} g^{\mu \nu} \partial_\mu \phi  \partial_\nu \phi + V(\phi) \right)\,, 
\eeq
where $\hat{R}$ is the conformally transformed Ricci scalar and the potential  has the form,
\beq
V(\phi) = \frac{k^2}{4} \frac{f(R) - R F(R)}{ F^2(R)}\,. 
\eeq
 For the sake of concreteness and to make our point even more transparent, we will consider the following ad-hoc gravity during inflation \footnote{Although every single inflationary model is a toy model, we would like to stress that there is not a physics motivation for the proposed modification of gravity. At the same time, we should also bear in mind that physics without assumptions or  caveats is unimaginable. We thus, left the reader judge by himself the degree of skepticism that is appropriate when considering modified gravity models of inflation like the one presented here.}, 
\beq
f(R) =  R \left( 1 + (R/M^2) ^{5/4} \right) \,,
\eeq
where $M$ is an arbitrary  mass scale and we assume that during inflation the second term dominates over the first one, implying that during inflation $H^2 \gg M^2$.  Clearly, once the inflationary phase is over, we smoothly approach Einstein gravity  
In this case,
\beq
F(R) = \exp{\left( \sqrt{\frac{2}{3}} \phi\right)}\,,
\eeq
and 
\beq
V(\phi) = \frac{-5 k^2}{54} \left(\frac{2}{3}\right)^{3/5} M^2 e^{-2 \sqrt{\frac{2}{3}} \phi} \left(e^{\sqrt{\frac{2}{3}} \phi}-1\right)^{9/5}\,.
\eeq
As before, the form of the potential looks rather complicated but its shape is pretty simple,
as can be seen from figure~\ref{fr}. In fact, already by eye, we can guess that this kind of potential
should be able to accommodate a decent period of inflation.
As seen in Model B, the cutoff scale in this kind of potentials is again $M_{Pl}$ and the effective theory is UV safe.

But we can do way better that guessing; the analysis of our potential is straightforward\footnote{Notice that unlike the previous cases, now we are not using the Hamilton-Jacobi formalism any longer and therefore we
need to resort to another set of slow-roll parameters, the ones calculated directly from the potential. Their relation with the spectral index takes into account this difference (see for instance \cite{Liddle:1994dx}).},
\bea
\epsilon_V &=& \frac{k^2}{2} \left( \frac{V'}{V} \right)^2 =
 \frac{1}{300} \left(11-9~ \coth \left(\frac{\phi}{\sqrt{6}}\right)\right)^2\,, \\
\eta_V&=& k^2 \frac{V''}{V} =
\frac{2}{75} \left(\frac{36}{\left(e^{\sqrt{\frac{2}{3}} \phi}-1\right)^2}-\frac{63}{e^{\sqrt{\frac{2}{3}} \phi}-1}+1\right)\,,\\
n_s&=&  1 - 6 \epsilon_V + 2 \eta_V =\frac{1}{150} \left(198 \coth \left(\frac{\phi}{\sqrt{6}}\right)-171 \csch^2\left(\frac{\phi}{\sqrt{6}}\right)-52\right)\,.
\eea
The modes we are interested in studying are those that left the horizon 60 efoldings before the end of inflation, where $\phi_{\rm end}$,  the field value at the end of inflation, is calculated by asking $\epsilon_V =1$. Then, the field value at horizon exit is obtained from, 
\beq
N = 62= \int_{\phi_{\mbox{\tiny end}}}^{\phi_{\mbox{\tiny hor}}}  5 \sqrt{\frac{3}{2}} \left(-\frac{9}{e^{\sqrt{\frac{2}{3}} \phi }-10}-1\right) d\phi \,,
\eeq
which is independent of the specific value of $M$ and, for our choice of parameters, is well beyond $M_{\rm Pl}$,  $\phi_{\rm hor} \approx 15 \;  M_{\rm Pl}$,  giving, 
\bea
n_s = 0.97, \qquad r= 0.16 \,.
\eea
Once again, we see that an innocent modification of gravity, when casted as a minimally coupled scalar field rolling down a potential, ends up giving transplanckian field values.
Once more, we would like to stress that, as in the previous example, we have not
designed a modification of gravity able to accommodate transplanckian field values. We
have just chosen an $f(R)$ capable of producing sizeable tensor modes and found that this corresponds
to transplanckian field values once analysed as a minimally coupled scalar field.

\section{Conclusions}
\label{sec:conclusions}
In this work, we have explored the possibility that the transplanckian field values needed to accommodate the experimental results in minimally coupled single-field inflation models are only due to our insistence of imposing a minimal coupling of the inflaton field to gravity. If the theory responsible for inflation 
includes a non-minimal
coupling to gravity, the energies and field values can be subplanckian
during the full inflation era in the Jordan frame, while they may
appear transplankian in the Einstein frame.

We are perfectly aware that the field value by itself carries no information, it is 
after all a ``dummy'' variable, but the fact that its vacuum expectation value turns out to be
well above the Planck mass may be telling us that it is gravity (or its couplings to gravity),
and not only the inflaton potential couplings, the true drivers of inflation.
 
We have shown (Section \ref{monomial}: Monomial potentials ) that not only it is possible to turn the most popular inflationary potentials
($\phi^4$, $\phi^2$) into the desired regime by choosing an appropriate coupling to gravity,
but also that scalar tensor theories, designed exclusively to sustain inflation by asking the 
scale factor to grow exponentially (Section \ref{generic}: Generic scalar-tensor theories), also turn subplanckian even in the simplest cases
(let us remind the reader that the case $H'/H = 0 $ was discarded, not because it does not
satisfy our conjecture, but because it leads to a spectral index larger than 1).
We have also presented a case (Section \ref{fR}: f(R) gravity models) where gravity itself is solely responsible for inflation,
and again results in transplanckian field values once interpreted as single-field inflation.

In summary, we have seen that is possible, and in fact quite natural
and easy, to find realistic examples of theories with non-minimal
coupling to gravity that have transplanckian field values if we insist
on imposing a minimal coupling to gravity, but are always subplanckian
in their ``natural'' frame. Thus, we have proven that single-field inflation models can
still accommodate a large tensor-to-scalar ratio with subplankian
field values in the presence of non-minimal coupling to gravity.

\section*{Acknowledgements}
The authors are grateful to Scott Dodelson and Joe Lykken
for useful discussions.  We acknowledge support from the MEC and
FEDER (EC) Grants
FPA2011-23596 and the Generalitat Valenciana under grant PROMETEOII/2013/017.
G.B. acknowledges partial support from the European Union FP7 ITN INVISIBLES (Marie
Curie Actions, PITN-GA-2011-289442).

\end{document}